\documentclass{appolb}
\usepackage{graphicx}
\usepackage{amsmath}
\usepackage{bbold}
\usepackage{amssymb}
\usepackage{esint}
\usepackage{enumitem}
\usepackage{multirow}
\usepackage{graphicx}
\usepackage{hhline}
\usepackage{hyperref}
\usepackage{centernot}
\usepackage{ragged2e}
\usepackage{amsthm}
\hypersetup{
  colorlinks   = true,    
  urlcolor     = blue,    
  linkcolor    = blue,    
  citecolor    = black     
}

\begin{document}
\title{Wilson line-based action for gluodynamics at the quantum level%
\thanks{Presented at \emph{XXX Cracow EPIPHANY Conference on Precision Physics at High Energy Colliders}}%
}
\author{Bartosz Grygielski 
\address{Jagiellonian University, 
S. Lojasiewicza 11, 30-348 Krak\'ow, Poland}
\\[3mm]
{Hiren Kakkad\thanks{Speaker},
 Piotr Kotko
\address{AGH University of Krakow, al. Mickiewicza 30, 30-059, Poland}
}
}
\maketitle
\begin{abstract}
We recently derived a new action for gluodynamics by canonically transforming the Yang-Mills action on light-cone. The transformation elimated triple gluons vertices and replaced the gauge fields with Wilson lines. This greatly reduced the number of diagrams required to compute tree level amplitudes. However, at the quantum level, the action turned out to be incomplete. We present two ways, based on one-loop effective action approach, to systematically develop quantum correction to our action. The first method retains Yang-Mills vertices in the loop, while the second method explicitly incorporates the interaction vertices of our action into the loop. We demonstrate that both approaches are equivalent, although the former appears to be more efficient for computing higher-multiplicity one-loop amplitudes.
\end{abstract}
  
\section{Introduction}
Scattering amplitudes are fundamental for predicting cross sections in particle colliders. Yet, computing them in theories with self-interactions, such as the Yang-Mills theory for gluons, is challenging. Traditionally, this computation relied on Feynman diagrams, but even at the classical level (tree amplitudes), the sheer number of diagrams can become overwhelming, despite some cases like the \textit{maximally helicity violating} (MHV) having relatively simple result \cite{Parke:1986gb}. Consequently, there is always a search for a new, more efficient approach for computing pure gluonic amplitudes.

Recent progress in this direction has predominantly concentrated on the so-called \textit{on-shell} approaches abandoning the more traditional space-time field theoretic action-based approach. On the contrary, we recently derived a new action -- we dub it the "Z-field" action -- by canonically transforming the Yang-Mills action on the light cone \cite{Scherk1975}. The latter can be represented schematically as follows: 
\begin{multline}
    S_{\mathrm{YM}}[A]
    = -A^{\star L}\square_{LJ}A^{\bullet J} -\left(V_{-++}^{IJK}\right)A^{\star I} A^{\bullet J}A^{\bullet K} \\
    -\left(V_{--+}^{KIJ}\right)A^{\star K} A^{\star I} A^{\bullet J}-\left(V_{--++}^{LIJK}\right)A^{\star L} A^{\star I} A^{\bullet J} A^{\bullet K}.
   \label{eq:S[A]}
\end{multline}
Throughout this manuscript, we predominantly employ collective indices $I, J, K, \dots$, which encompass color and position associated with the fields. Repeated indices are summed over. However, if necessary, we can revert to the more conventional notation with explicit color and position. For instance, $A^{\star L} \equiv \hat{A}^{\star}(x)=A_a^{\star}t^a$, where $t^a$ are the color generators satisfying $\left[t^{a},t^{b}\right]=i\sqrt{2}f^{abc}t^{c}$. For position, we employ the 'double-null' coordinates defined as $v^{+}=v\cdot\eta$, $v^{-}=v\cdot\tilde{\eta}$, $v^{\bullet}=v\cdot\varepsilon_{\bot}^{+}$, and $v^{\star}=v\cdot\varepsilon_{\bot}^{-}$, with $\eta=\left(1,0,0,-1\right)/\sqrt{2}$, $\tilde{\eta}=\left(1,0,0,1\right)/\sqrt{2}$, and $\varepsilon_{\perp}^{\pm}=\frac{1}{\sqrt{2}}\left(0,1,\pm i,0\right)$. In these coordinates, $\square=2(\partial_+\partial_- - \partial_{\bullet}\partial_{\star})$.

The light-cone Yang-Mills action Eq.~\eqref{eq:S[A]} relies solely on the two transverse field components $A^{\bullet}=(A^{1}+iA^{2})/\sqrt{2}$ and $A^{\star}=(A^{1}-iA^{2})/\sqrt{2}$, which, in the on-shell limit, can be identified with the 'plus' and 'minus' helicities. In our convention, $\bullet =$ plus and $\star =$ minus. 

It is well-known that the triple gluon vertices -- $V_{--+}$ and $V_{-++}$ in Eq.~\eqref{eq:S[A]} -- vanish for real momenta in the on-shell limit. Moreover they are rather small "building blocks" due to which the number of Feynman diagrams 
 grows factorially when computing higher multiplicity amplitude. We therefore decided to derive a new action by performing a canonical transformation that eliminates these vertices from the Yang-Mills action Eq.~\eqref{eq:S[A]} \cite{Kakkad:2021uhv}. The solution of the transformation has the following form
\begin{equation}
A^{\star L}[Z] = 
\sum_{n=1}^{\infty} 
     \, \sum_{i=1}^{n}\,\Lambda_{i,n-i}^{L \{J_1 \dots J_i\}\{ J_{i+1}\dots J_n\} } \,\prod_{k=1}^{i}{Z}^{\star J_k} \prod_{l=i+1}^{n}{Z}^{\bullet J_l}\,,
    \label{eq:A_star_Za}
\end{equation}
and
\begin{equation}
A^{\bullet L}[Z] = 
\sum_{n=1}^{\infty} 
     \, \sum_{i=1}^{n}\, \Xi_{i,n-i}^{L \{J_1 \dots J_i\}\{ J_{i+1}\dots J_n\} }\,\prod_{k=1}^{i}{Z}^{\bullet J_k} \prod_{l=i+1}^{n}{Z}^{\star J_l}\,.
    \label{eq:A_bul_Za}
\end{equation}
 Substituting Eqs.~\eqref{eq:A_star_Za}-\eqref{eq:A_bul_Za} in the Yang-Mills action Eq.~\eqref{eq:S[A]} we get the new action which has the following form
\begin{equation}
    S[Z]
    = -Z^{\star L}\square_{LJ}Z^{\bullet J} -\sum_{n=4}^{\infty} 
     \, \sum_{m=2}^{n-2}\,
    \,\, \mathcal{U}^{ \{J_1 \dots J_m\}\{ J_{m+1}\dots J_n\} }_{\underbrace{-\,\cdots\,-}_{m}\underbrace{+ \,\cdots\, +}_{n-m}} 
   \prod_{i=1}^{m}Z^{\star J_i}
   \prod_{k=m+1}^{n}Z^{\bullet J_k} \, .
   \label{eq:S[Z]}
\end{equation}
For our current discussion, the exact expressions for the kernels $\Lambda_{i,j}$ and $\Xi_{i,j}$ as well as the interaction vertices $\mathcal{U}^{ \{J_1 \dots J_m\}\{ J_{m+1}\dots J_n\} }_{-\,\cdots\,- + \,\cdots\, +}$ are not required. These can, however, be found in \cite{Kakkad:2021uhv, kakkad2023scattering}. Notice, starting from 4-point MHV there are higher point interaction vertices in the Z-field action. This greatly reduces the number of diagrams required to compute tree level amplitudes. For instance, 9-point amplitudes require a maximum of 25 diagrams \cite{Kulig2024}. Furthermore, the number of diagrams required to compute \textit{split-helicity} tree level amplitudes using the Z-field action follows \textit{Delannoy numbers} \cite{kakkad2023scattering}. Another interesting aspect is that the inverse of the solutions Eqs.~\eqref{eq:A_star_Za}-\eqref{eq:A_bul_Za} have a 3-dimensional structure of intersecting lines. For details and an elaborate discussion for tree level computation see\footnote{A concise presentation can also be found in "Computing multi-leg scattering amplitudes using light-cone actions" in these proceedings.} \cite{Kotko2017, Kakkad2020,Kakkad:2021uhv, kakkad2023scattering}.

In the present text, our focus is rather on discussing quantum corrections to the Z-field action Eq.~\eqref{eq:S[Z]}. Notice, amplitudes of the type $(+ \dots + \pm)$ and $(- \dots - \pm)$ are all zero in the Z-field action. While these amplitudes are zero at tree level, they become non-zero at one-loop level, suggesting the absence of certain loop contributions in Z-field action Eq.~\eqref{eq:S[Z]}. Below, we delve into methods to address this issue by systematically developing loop corrections to the Z-field action.

\section{Quantum corrections}
Below we present two \textit{equivalent} methods for systematically incorporating quantum corrections to the Z-field action Eq.~\eqref{eq:S[Z]}. Both rely on the one-loop effective action method. We explore the strengths and weaknesses of each approach and demonstrate their equivalence.
\subsection{Approach 1: Yang-Mills vertices in the loop}

In this approach, we start with the generating functional for the full Greens function for the Yang-Mills theory (for the sake of simplicity we make color and position explicit)
\begin{equation}
    \mathcal{Z}[J]=\int[dA]\, e^{i\left(S_{\mathrm{YM}}[A] + \int\!d^4x\, \Tr \hat{J}_j(x) \hat{A}^j(x)\right) } \,,
    \label{eq:gen_YM}
\end{equation}
where $S_{\mathrm{YM}}$ is the light-cone Yang-Mills action Eq.~\eqref{eq:S[A]}. $\hat{J}=J_a t^a$ is the auxiliary current coupled with the Yang-Mills fields and the index $j = \bullet, \star$ runs over the transverse field components. To derive the one-loop effective action, we expand the terms in the exponent around the classical configuration ${\hat A}^i_{c}=\left\{{\hat A}_{c}^{\bullet}(x), {\hat A}_{c}^{\star}(x) \right\}$ up to second order in fields. Higher-order terms are necessary for corrections beyond one loop. The linear term vanishes owing to the classical EOMs. Performing the Gaussian integral we get \cite{Kakkad_2022,kakkad2023scattering}
\begin{equation}
    \mathcal{Z}[J]\approx 
    e^{ iS_{\mathrm{YM}}[A_c[J]] 
    + i\int d^4x \Tr \hat{J}_i(x) \hat{A}_c^i[J](x)  -\frac{1}{2} \Tr\ln \left[ \mathrm{M}^{\mathrm{YM}}[J]\right]
    }\,.
\label{eq:Partition_YM}
\end{equation}
where $\hat{A}_c^i(x)$ is understood as functional of the currents $\hat{A}_c^i[J](x)$ obtained from solving the classical EOMs and the matrix in the log term reads
\begin{equation}
\mathrm{M}^{\mathrm{YM}}_{IN}[J]= \left(\begin{matrix}
     \frac{\delta^2 S_{\mathrm{YM}}[A_c]}
    {\delta A^{\bullet I}\delta A^{\star N}} 
     & \frac{\delta^2 S_{\mathrm{YM}}[A_c]}
    {\delta A^{\bullet I}\delta A^{\bullet N}} \\ \\
\frac{\delta^2 S_{\mathrm{YM}}[A_c]}
    {\delta A^{\star I}\delta A^{\star N}} & \frac{\delta^2 S_{\mathrm{YM}}[A_c]}
    {\delta A^{\star I}\delta A^{\bullet N}}    
\end{matrix}\right)\,.
\label{eq:MAT}
\end{equation}
The one-loop effective action $\Gamma_{\mathrm{YM}}[A_c]$ is defined as the Legendre transform of the generating functional for the connected Greens function
\begin{equation}
   \Gamma_{\mathrm{YM}}[A_c] = W_{\mathrm{YM}}[J] - \int\!d^4x\, \Tr \hat{J}_i(x) \hat{A}_c^i(x) \,,
\end{equation}
where $W_{\mathrm{YM}}[J] = -i \ln \left[ \mathcal{Z}_{\mathrm{YM}}[J]\right]$. Substituting Eq.~\eqref{eq:Partition_YM} we get
\begin{multline}
   \Gamma_{\mathrm{YM}}[A_c] = S_{\mathrm{YM}}[A_c] 
    + \frac{i}{2} \Tr\ln \Bigg[ \frac{\delta^2 S_{\mathrm{YM}}[A_c]}
    {\delta A^{\star I}\delta A^{\bullet K}} \, \frac{\delta^2 S_{\mathrm{YM}}[A_c]}
    {\delta A^{\star K}\delta A^{\bullet J}} \\
    - \frac{\delta^2 S_{\mathrm{YM}}[A_c]}
    {\delta A^{\star I}\delta A^{\bullet K}} \, \frac{\delta^2 S_{\mathrm{YM}}[A_c]}
    {\delta A^{\star K}\delta A^{\star L}} \left( \frac{\delta^2 S_{\mathrm{YM}}[A_c]}
    {\delta A^{\bullet L}\delta A^{\star M}} \right)^{-1} \frac{\delta^2 S_{\mathrm{YM}}[A_c]}
    {\delta A^{\bullet M}\delta A^{\bullet J}}\Bigg]\,,
    \label{eq:OLEA_YM}
\end{multline}
where we used the identity Eq.~\eqref{eq:lnM_decomp} in the log term. The above expression depends only on fields and represents the Yang-Mills one-loop effective action where the log term accounts for the loop contribution. To see this, notice that each of the two terms in the log consists of second-order derivative of Yang-Mills action with respect to $A^{\star} $ and $ A^{\bullet}$. At the lowest order in fields this term gives an inverse propagator which can be factored out of the log as follows

\begin{multline}
    \Tr\ln \left[ \mathrm{M}^{\mathrm{YM}}\right] = \Tr\ln [-\square] \\
    + \Tr\ln \Bigg[  \Bigg\{ \Big( \mathbb{1} + \left(\square^{-1}V_{-++}^{IKP}\right)A_{c}^{\bullet P} +\left(\square^{-1}V_{--+}^{PIK}\right)A_{c}^{\star P} +\left(\square^{-1}V_{--++}^{PIKQ}\right)A_{c}^{\star P} A_{c}^{\bullet Q}\Big) \nonumber
    \end{multline}
    \begin{multline}
    \times \Big( \mathbb{1} + \left(\square^{-1}V_{-++}^{KJR}\right)A_{c}^{\bullet R} +\left(\square^{-1}V_{--+}^{RKJ}\right)A_{c}^{\star R} +\left(\square^{-1}V_{--++}^{RKJS}\right)A_{c}^{\star R} A_{c}^{\bullet S}\Big) \Bigg\}\\
     - \Bigg\{ \Big(\mathbb{1} + \left(\square^{-1}V_{-++}^{IKP}\right)A_{c}^{\bullet P} +\left(\square^{-1}V_{--+}^{PIK}\right)A_{c}^{\star P} +\left(\square^{-1}V_{--++}^{PIKQ}\right)A_{c}^{\star P} A_{c}^{\bullet Q}\Big) \nonumber
     \end{multline}
    \begin{multline} \times \Big(\left(\square^{-1}V_{--+}^{KLR}\right)A_{c}^{\bullet R} +\left(\square^{-1}V_{--++}^{KLRS}\right)A_{c}^{\bullet R} A_{c}^{\bullet S}\Big)\\
     \times \Big( \mathbb{1} + \left(\square^{-1}V_{-++}^{MLT}\right)A_{c}^{\bullet T} +\left(\square^{-1}V_{--+}^{TML}\right)A_{c}^{\star T} +\left(\square^{-1}V_{--++}^{TMLU}\right)A_{c}^{\star T} A_{c}^{\bullet U}\Big)^{-1}\\ 
     \times \Big(\left(\square^{-1}V_{-++}^{VMJ}\right)A_{c}^{\star V} +\left(\square^{-1}V_{--++}^{WVMJ}\right)A_{c}^{\star W} A_{c}^{\star V}\Big)\Bigg\}\Bigg]
    \label{eq:Partition_log}
    \,.
\end{multline}
The first log is field-independent and can be discarded. However, factoring it out equips the other terms with a propagator necessary to connect vertices as is well-known from the Feynman rules. The second log can be expanded into a series. 
Doing this and finally tracing over the differentiated legs generates the loops. Following this procedure, up to second-order in fields, the log term  reads
\begin{multline}
  \Tr\ln \left[ \mathrm{M}^{\mathrm{YM}}[A]\right] \Bigg|_{2nd}= 2\, \left(\square^{-1}V_{-++}^{IIP}\right) A_c^{\bullet P} +2\, \left(\square^{-1}V_{--+}^{PII}\right) A_c^{\star P} \\
  +2\, \left(\square^{-1}V_{--++}^{PIIQ}\right) A_c^{\star P}A_c^{\bullet Q} 
    -\left(\square^{-1}V_{-++}^{IKP}\square^{-1}V_{-++}^{KIR}\right) 
    A_c^{\bullet P} A_c^{\bullet R}\\
    -\left(\square^{-1}V_{--+}^{PIK}\square^{-1}V_{--+}^{RKI}\right) 
    A_c^{\star P} A_c^{\star R}
    -\Big[2\,\left(\square^{-1}V_{-++}^{IKP}\square^{-1}V_{--+}^{RKI}\right) 
    \\
    +\,\left(\square^{-1}V_{--+}^{IMP}\square^{-1}V_{-++}^{RMI}\right)\Big] 
    A_c^{\bullet P} A_c^{\star R}  \,,
    \label{eq:2p_1loopexp}
\end{multline}
where the first three terms are tadpoles and the remaining three are bubbles.

To develop one-loop corrections to the Z-field action, we performed the transformation \cite{Kakkad:2021uhv} that derives the Z-field action from the Yang-Mills action to the Yang-Mills one-loop effective action Eq.~\eqref{eq:OLEA_YM}. By doing this we get
\begin{figure}
    \centering
    \parbox[c]{2.4cm}{ \includegraphics[width=2.4cm]{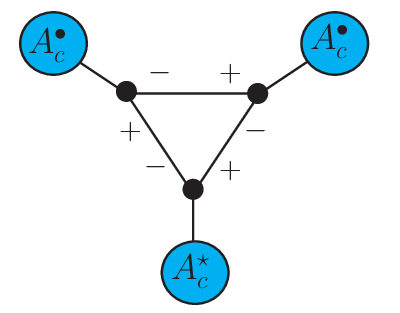}}\qquad $\xrightarrow{\hat{A}_c^i(x) \Longrightarrow \hat{A}_c^i[Z_c](x)}$ \qquad\qquad
    \parbox[c]{4.8cm}{
    \includegraphics[width=4.8cm]{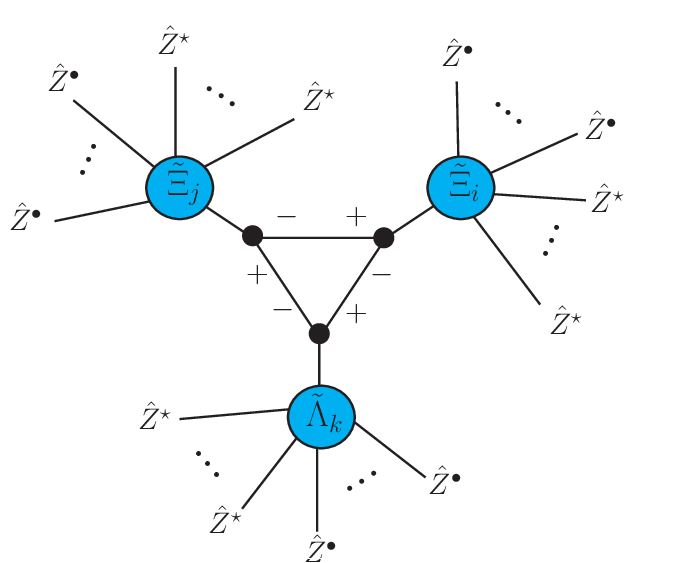}}
    \caption{\small
    On the left, we have a $(+ + -)$ one-loop triangular contribution originating from the log term in the Yang-Mills one-loop effective action Eq.~\eqref{eq:OLEA_YM}. After performing the canonical transformation, the field outside the triangle undergoes the substitution $\hat{A}_c^i(x) \rightarrow \hat{A}_c^i[Z_c](x)$. This, however, does not alter the loop itself. 
    }
    \label{fig:zloop_st}
\end{figure}

\begin{multline}
   \Gamma[Z_c] = S[Z_c] 
    + \frac{i}{2} \Tr\ln \left[ \frac{\delta^2 S_{\mathrm{YM}}[A]}
    {\delta A^{\star I}\delta A^{\bullet K}} \, \frac{\delta^2 S_{\mathrm{YM}}[A]}
    {\delta A^{\star K}\delta A^{\bullet J}} \right. \\
    \left.- \frac{\delta^2 S_{\mathrm{YM}}[A]}
    {\delta A^{\star I}\delta A^{\bullet K}} \, \frac{\delta^2 S_{\mathrm{YM}}[A]}
    {\delta A^{\star K}\delta A^{\star L}} \left( \frac{\delta^2 S_{\mathrm{YM}}[A]}
    {\delta A^{\bullet L}\delta A^{\star M}} \right)^{-1} \frac{\delta^2 S_{\mathrm{YM}}[A]}
    {\delta A^{\bullet M}\delta A^{\bullet J}}\right]_{A_c=A_c[Z_c]}\,.
    \label{eq:OLEA_Z}
\end{multline}
The transformation maps $S_{\mathrm{YM}}[A_c] \rightarrow S[Z_c]$ and for the log term it substitutes the Yang-Mills fields outside the loop (all the $\hat{A}_c^i(x) $ fields in Eq.~\eqref{eq:Partition_log}) with the solution $\hat{A}_c^i[Z_c](x)$ of the transformation Eqs.~\eqref{eq:A_star_Za}-\eqref{eq:A_bul_Za}. As a result $\mathrm{M}^{\mathrm{YM}}[A]$ in Eq.~\eqref{eq:OLEA_YM} becomes $\mathrm{M}^{\mathrm{YM}}[A[Z]] \equiv \mathrm{M}^{\mathrm{YM}}[Z]$ . Following the same procedure discussed above for expanding the log term in Eq.~\eqref{eq:OLEA_YM}, up to the second order in Z-fields the log term in Eq.~\eqref{eq:OLEA_Z} reads
\begin{multline}
  \Tr\ln \left[ \mathrm{M}^{\mathrm{YM}}[Z]\right] \Bigg|_{2nd}= 2\, \left(\square^{-1}V_{-++}^{IIP}\right) Z_c^{\bullet P} +2\, \left(\square^{-1}V_{--+}^{PII}\right) Z_c^{\star P} \\
   +\Big[2\,\left(\square^{-1}V_{-++}^{IIP}\right) \Xi_{2,0}^{P \{J_1 J_2\}} -\left(\square^{-1}V_{-++}^{IKJ_1}\square^{-1}V_{-++}^{KIJ_2}\right) \Big]
    Z_c^{\bullet J_1} Z_c^{\bullet J_2}\\
   + \Big[2\, \left(\square^{-1}V_{--+}^{PII}\right) \Lambda_{2,0}^{P \{J_1 J_2\} } -\left(\square^{-1}V_{--+}^{J_1IK}\square^{-1}V_{--+}^{J_2KI}\right) \Big]
    Z_c^{\star J_1} Z_c^{\star J_2}
  \\
  + \Big[2 \left(\square^{-1}V_{-++}^{IIP}\right) \Xi_{1,1}^{P \{J_2\}\{ J_{1}\} }
  +2 \left(\square^{-1}V_{--+}^{PII}\right) \Lambda_{1,1}^{P \{J_1 \}\{ J_{2}\} } 
  +2 \left(\square^{-1}V_{--++}^{J_1IIJ_2}\right) \\
  -2\left(\square^{-1}V_{-++}^{IKJ_2}\square^{-1}V_{--+}^{J_1KI}\right) 
    -\left(\square^{-1}V_{--+}^{IMJ_2}\square^{-1}V_{-++}^{J_1MI}\right)\Big] \,{Z}_c^{\star J_1} {Z}_c^{\bullet J_2} .
    \label{eq:2p_1loopexpZth}
\end{multline}
Note, the above expression represents all the one-loop contributions necessary to compute amputated one-loop Green's function up to two points. Whereas if one wants to compute the same object using the Yang-Mills one-loop effective action Eq.~\eqref{eq:OLEA_YM}, the terms shown in Eq.~\eqref{eq:2p_1loopexp} are incomplete because one can take the tadpoles from the latter and connect it with the triple gluons vertices from $S_{\mathrm{YM}}[A_c]$ to obtain additional contributions. In the case of Z-fields one-loop effective action Eq.~\eqref{eq:OLEA_Z}, the substitution $\hat{A}_c^i(x) \rightarrow \hat{A}_c^i[Z_c](x)$ in the log term accounts for these. This idea generalises as follows. The substitution $\hat{A}_c^i(x) \rightarrow \hat{A}_c^i[Z_c](x)$ outside the loops in Eq.~\eqref{eq:OLEA_Z} accounts for all the tree-level connections involving the triple gluons vertices which in the case of Yang-Mills one has to do separately. This combined with the fact that the interactions vertices in our action are much "bigger" than the Yang-Mills vertices implies that computing higher multiplicity one-loop amplitudes using Eq.~\eqref{eq:OLEA_Z} should be much more efficient.

Although by construction the action Eq.~\eqref{eq:OLEA_Z} should be one-loop complete with no missing contributions, we validated it by computing 4-point one amplitudes with all helicity combinations: $(+ + + \pm)$, $(+ + - -)$, and $(- - - \pm)$ in \cite{kakkad2023scattering}. We also used this approach to successfully develop quantum corrections to the MHV action \cite{Mansfield2006} (understood as the action implementing the CSW rules \cite{Cachazo2004}) which suffers from similar issues \cite{Kakkad_2022}. 

The only drawback of this approach is that the substitution $\hat{A}_c^i(x) \rightarrow \hat{A}_c^i[Z_c](x)$ in the log term of Eq.~\eqref{eq:OLEA_Z} does not affect the loop structure (see Figure \ref{fig:zloop_st}). That is, the Yang-Mills vertices remain intact in the loop whereas the Z-field interaction vertices are only available for tree-level connections outside the loops. We would therefore like to re-derive the Z-field one-loop effective action such that the Z-field vertices are explicit in the loop.
\subsection{Approach 2: Z-field vertices in the loop}

In this approach, the starting point is still the generating functional for full Green's function for the Yang-Mills theory Eq.~\eqref{eq:gen_YM}. The change is we perform the canonical transformation that derives the Z-field action from the Yang-Mills action right away to get
\begin{equation}
    \mathcal{Z}[J]=\int[dZ]\, e^{i\left(S[Z] + \int\!d^4x\, \Tr \hat{J}_j(x) \hat{A}^j[Z](x)\right) } \,.
   \label{eq:SZ_genr}
\end{equation}
The important point is to also transform the source term in Eq.~\eqref{eq:gen_YM} which now becomes non-linear in fields. As a result when we repeat the derivation of one-loop approximation by expanding around the classical configuration up to the second order in fields followed by the Gaussian integration, we get
\begin{equation}
   \mathcal{Z}[J] \approx  
   e^{iS\left[Z_c\left[J\right]\right] 
    + i\int\!d^4x\, \Tr \hat{J}_i(x)\hat{A}^i[Z_c[J]](x) -\frac{1}{2}\Tr\ln\mathrm{M}[J]} \,.
    \label{eq:det_SZln}
\end{equation}
In this case, the matrix in the log is a bit more complicated
\begin{equation}
  \mathrm{M}[J]=\mathrm{M}^{\text{Z-field}}[J]+\mathrm{M}^{\text{src}}[J] \,,  
  \label{eq:Mfull}
\end{equation}
where
 \begin{equation}
\mathrm{M}^{\text{Z-field}}_{IK}[J]  =  \left(\begin{matrix}
     \frac{\delta^2 S[Z_c]}{\delta Z^{\bullet I}\delta Z^{\star K}} 
     &
     \frac{\delta^2 S[Z_c]}{\delta Z^{\bullet I}\delta Z^{\bullet K}}\\ \\
\frac{\delta^2 S[Z_c]}{\delta Z^{\star I}\delta Z^{\star K}}
      &
      \frac{\delta^2 S[Z_c]}{\delta Z^{\star I}\delta Z^{\bullet K}} 
\end{matrix}\right) \,,
\label{eq:M_Zfield}
\end{equation}
and
\begin{equation}
\mathrm{M}^{\text{src}}_{IK}[J] = \left(\begin{matrix}
    J_{\star L}\frac{\delta^2 A^{\star L}[Z_c]}{\delta Z^{\bullet I}\delta Z^{\star K}} + J_{\bullet L}\frac{\delta^2 A^{\bullet L}[Z_c]}{\delta Z^{\bullet I}\delta Z^{\star K}} 
     & J_{\star L}\frac{\delta^2 A^{\star L}[Z_c]}{\delta Z^{\bullet I}\delta Z^{\bullet K}} +  J_{\bullet L}\frac{\delta^2 A^{\bullet L}[Z_c]}{\delta Z^{\bullet I}\delta Z^{\bullet K}}\\ \\
 J_{\star L}\frac{\delta^2 A^{\star L}[Z_c]}{\delta Z^{\star I}\delta Z^{\star K}} + J_{\bullet L}\frac{\delta^2 A^{\bullet L}[Z_c]}{\delta Z^{\star I}\delta Z^{\star K}}
      &J_{\star L}\frac{\delta^2 A^{\star L}[Z_c]}{\delta Z^{\star I}\delta Z^{\bullet K}} + J_{\bullet L}\frac{\delta^2 A^{\bullet L}[Z_c]}{\delta Z^{\star I}\delta Z^{\bullet K}}
\end{matrix}\right) \,.
\label{eq:M_Zsrc}
\end{equation}
The matrix $\mathrm{M}^{\text{Z-field}}[J] $ in Eq.~\eqref{eq:M_Zfield}  accounts for all the one-loop contributions involving only the Z-field interaction vertices. Since we know that Z-field vertices do not give the full one-loop amplitudes, it is naturally expected that there must be additional contributions. These as we demonstrate below are accounted by the "source-matrix" $ \mathrm{M}^{\text{src}}[J]$ Eq.~\eqref{eq:M_Zsrc}. Notice, instead of transforming the source term in Eq.~\eqref{eq:gen_YM}, had we coupled a "new" source term of the type $\int\!d^4x\hat{J}'_i(x)\hat{Z}_i(x)$ with the Z-filed action then repeating the above derivation, we would only get the matrix $\mathrm{M}^{\text{Z-field}}[J] $ in Eq.~\eqref{eq:M_Zfield}. 

\begin{figure}
    \centering
    \includegraphics[width=11cm]{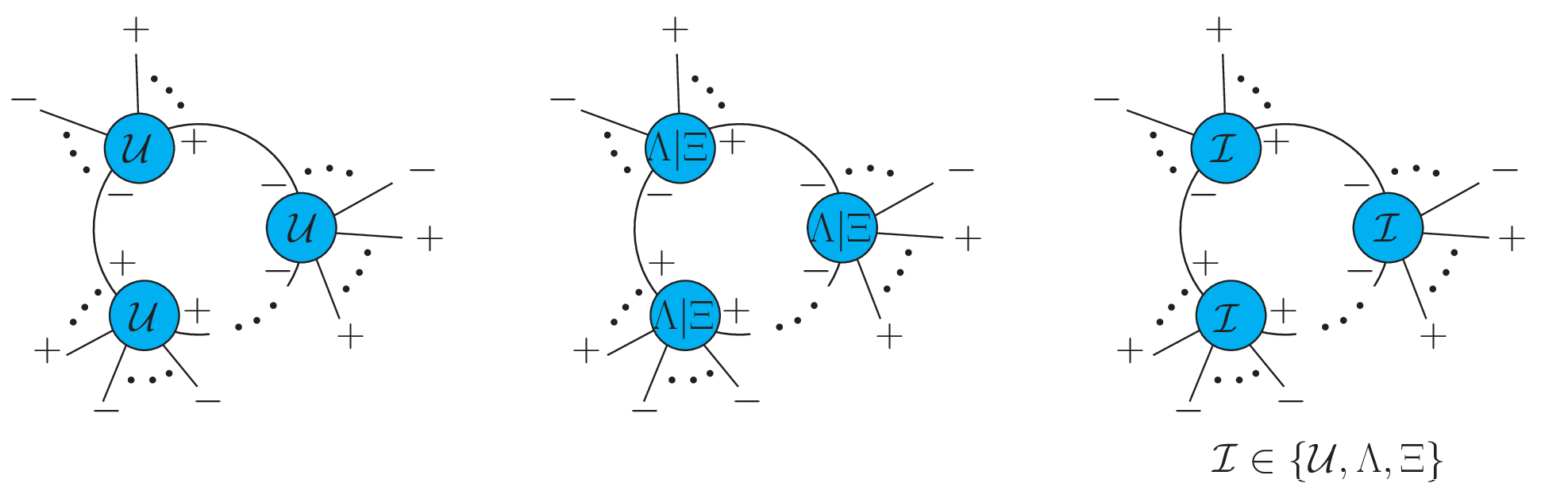}
    \caption{
    \small 
  The loop diagrams originating from the log term in Eq.~\eqref{eq:olea_zth} can be categorized into three sets: first includes those involving only Z-field vertices $\mathcal{U}$ in the loop, second includes those originating from the source-matrix Eq.~\eqref{eq:M_Zsrc} alone and thus have only the kernels $\Lambda_{i,j}$ and $\Xi_{i,j}$ of the solutions Eqs.~\eqref{eq:A_star_Za}-\eqref{eq:A_bul_Za} in the loop, and finally the third type includes those that mix the previous two.}
    \label{fig:Zth_loop}
\end{figure}

Using Eq.~\eqref{eq:det_SZln}, we can derive the one-loop effective action as follows
\begin{equation}
   \Gamma[Z_c] = W[J] - \int\!d^4x\, \Tr \hat{J}_i(x) \hat{A}^{i}_c[Z_c](x) \,,
   \label{eq:gam_zn}
\end{equation}
where $W[J] = -i \ln \left[ \mathcal{Z}[J]\right]$. However, in the current approach, there is one more step. Recall, $\Gamma[Z_c]$ is a functional only of fields whereas the above procedure keeps the sources in the matrix Eq.~\eqref{eq:M_Zsrc} unaltered. We therefore need a way to replace the sources in the latter in terms of fields. This is achieved via the classical EOMs. In \cite{kakkad2023scattering}, we showed that the complicated classical EOMs obtained from Eq.~\eqref{eq:SZ_genr} can be re-expressed in terms of the Yang-Mills classical EOMs as follows
\begin{equation}
  \frac{\delta S_{\mathrm{YM}}[A[Z_c]]}{\delta A^{\star L}}= -J_{\star L}[Z_c]\,, \quad\quad
    \frac{\delta S_{\mathrm{YM}}[A[Z_c]]}{\delta A^{\bullet L}}= - J_{\bullet L}[Z_c] \,.
    \label{eq:J_currZ}
\end{equation}
Substituting the above in Eq.~\eqref{eq:M_Zsrc}, we get
\begin{equation}
   \Gamma[Z_c] = S\left[Z_c\right] 
    + i\frac{1}{2}\Tr\ln\mathrm{M}[J[Z_c]]\,.
    \label{eq:olea_zth}
\end{equation}
The trace of log can be expressed as follows
\begin{equation}
    \Tr\ln\mathrm{M}= \Tr\ln\mathrm{M}_{\bullet\star} +
    \Tr\ln\left(\mathrm{M}_{\star\bullet}
    - \mathrm{M}_{\star\star}\mathrm{M}^{-1}_{\bullet\star}\mathrm{M}_{\bullet\bullet}\right) \,,
    \label{eq:lnM_decomp}
\end{equation}
where $\mathrm{M}_{\bullet\star}=\mathrm{M}_{\star\bullet}$ are the diagonal blocks of the matrix Eq.~\eqref{eq:Mfull}, $\mathrm{M}_{\star\star}$ is the bottom-left and $\mathrm{M}_{\bullet\bullet}$ is the top-right block. We can then factor out the inverse propagator followed by expanding the log as a series. Finally tracing over the differentiated legs, we get the loops. These loop diagrams are of three types: those involving only Z-field vertices in the loop, those originating from the source-matrix Eq.~\eqref{eq:M_Zsrc} alone and thus have only the kernels $\Lambda_{i,j}$ and $\Xi_{i,j}$ of the solutions Eqs.~\eqref{eq:A_star_Za}-\eqref{eq:A_bul_Za} in the loop, and finally those that mix the previous two. We present these types in Figure \ref{fig:Zth_loop}. 

Expanding the log term in Eq.~\eqref{eq:olea_zth} up to second order in fields we get
\begin{multline}
      \Tr\ln\mathrm{M}\Bigg|_{2nd} = -2\frac{\Lambda_{1,1}^{L \{I\}\{ P\}} }{\square_{IP}} \square_{LJ}{Z}_c^{\bullet J}  -2\square_{LJ}\frac{\Xi_{1,1}^{L \{P\}\{ I\} } }{\square_{IP}}Z_c^{\star J}  \\
    - \Bigg( 4 \frac{\Lambda_{1,2}^{L \{I\}\{ P J_1\}} }{\square_{IP}}\square_{LJ_2} + 2\,\frac{\Lambda_{1,1}^{L \{I\}\{ P\}} }{\square_{IP}}\square_{LJ} \Xi_{2,0}^{J \{J_1 J_2\} } + 2\,\frac{\Lambda_{1,1}^{L \{I\}\{ P\}} }{\square_{IP}}\left(V_{-++}^{LJ_1 J_2}\right) \\ + \frac{\Lambda_{1,1}^{L_1 \{K_1\}\{ P\}} }{\square_{IP}} \square_{L_1 J_1} \frac{\Lambda_{1,1}^{L_2 \{I\}\{ P_1\}} }{\square_{K_1 P_1}} \square_{L_2 J_2}\Bigg){Z}_c^{\bullet J_1}{Z}_c^{\bullet J_2}\\
    -\Bigg( 2\,\frac{\Xi_{1,1}^{L \{P\}\{ I\} } }{\square_{IP}}\square_{LJ} \Lambda_{2,0}^{J \{J_1 J_2\} } +2\,\frac{\Xi_{1,1}^{L \{P\}\{ I\} } }{\square_{IP}}\left(V_{--+}^{J_1 J_2 L}\right) + 4\,\frac{\Xi_{1,2}^{L \{P\}\{ I J_1 \} } }{\square_{IP}}\square_{LJ_2} \\
    +  \square_{L_1 J_1}\frac{\Xi_{1,1}^{L_1 \{P\}\{ K_1\} } }{\square_{IP}}\square_{L_2 J_2}\frac{\Xi_{1,1}^{L_2 \{P_1\}\{ I\} } }{\square_{K_1P_1}}\Bigg){Z}_c^{\star J_1} {Z}_c^{\star J_2} \\
    + \Bigg(  8  \frac{\mathcal{U}^{ \{I  J_1\}\{P J_2\} }_{-\,- + \, +}}{\square_{IP}} \,
   -4\,\frac{\Xi_{2,0}^{L_1 \{PK_1\}  }}{\square_{IP}}\square_{L_1 J_1} \, \frac{\Lambda_{2,0}^{L_2 \{P_1I\} }}{\square_{K_1P_1}}\square_{L_2 J_2} - 4\, \frac{\Lambda_{2,1}^{L \{I J_1\}\{ P \}} }{\square_{IP}}\square_{LJ_2} \\
   - 2\,\frac{\Lambda_{1,1}^{L \{I\}\{ P\}} }{\square_{IP}} \square_{LJ} \Xi_{1,1}^{J \{J_2 \}\{J_1 \} } 
   - 2\,\frac{\Lambda_{1,1}^{L \{I\}\{ P\}} }{\square_{IP}} 2\left(V_{--+}^{LJ_1 J_2 }\right) - 4\,\frac{\Xi_{2,1}^{L \{P J_2\}\{ I  \} } }{\square_{IP}} \square_{LJ_1}\\
   - 2\,\frac{\Xi_{1,1}^{L \{P\}\{ I\} } }{\square_{IP}} \square_{LJ} \Lambda_{1,1}^{J \{J_1\}\{  J_2\} } - \frac{\Lambda_{1,1}^{L_1 \{K_1\}\{ P\}} }{\square_{IP}} \square_{L_1 J_2 } \square_{L_2 J_1}\frac{\Xi_{1,1}^{L_2 \{P_1\}\{ I\} } }{\square_{K_1 P_1}}\\
   - 2 \frac{\Xi_{1,1}^{L \{P\}\{ I\} } }{\square_{IP}} 2\left(V_{-++}^{J_1 L J_2}\right)  
    -\square_{L_1 J_1}\frac{\Xi_{1,1}^{L_1 \{P\}\{ K_1\} } }{\square_{IP}}\frac{\Lambda_{1,1}^{L_2 \{I\}\{ P_1\}} }{\square_{K_1P_1}} \square_{L_2 J_2 } \Bigg) Z_c^{\star J_1}{Z}_c^{\bullet J_2}.
    \label{eq:Log2_2point}
    \end{multline}
The above expression demonstrates the major drawback of this approach. Notice, on comparing it with the Eq.~\eqref{eq:2p_1loopexpZth} which represents all the one-loop contributions up to 2 points in the previous approach, we see that except for the first two types of tadpoles, the number of contributing terms for all the other cases got doubled. However, in \cite{kakkad2023scattering}, we demonstrated that both the one-loop actions action Eq.~\eqref{eq:OLEA_Z} and \eqref{eq:olea_zth} are equivalent.

Precisely, we showed that one can start with the complicated log term in Eq.~\eqref{eq:olea_zth} and just use the properties of the canonical transformation to reduce it to the log term in Eq.~\eqref{eq:OLEA_Z} modulo a field-independent volume-divergent factor which does not contribute to amplitude computation. Since the classical action $S[Z]$ is the same in both the one-loop actions Eq.~\eqref{eq:OLEA_Z} and \eqref{eq:olea_zth}, and the log terms can be interchanged, the two one-loop actions are equivalent. This equivalence not only validates the quantum completeness of Eq.~\eqref{eq:olea_zth} but also verifies our previous claim that the missing contributions indeed come from the source matrix Eq.~\eqref{eq:M_Zsrc}. However, it states that in rewriting the log term of Eq.~\eqref{eq:OLEA_Z} to make Z-field vertices explicit in the loop, we are expanding a compact expression into a plethora of terms.

From the discussion above, it appears that explicitly introducing Z-field interaction vertices into the loop doesn't improve efficiency. Rather, it probably complicates amplitude computation and introduces spurious contributions, which are primarily tadpole contributions. To validate this statement we employed both actions Eq.~\eqref{eq:OLEA_Z} and \eqref{eq:olea_zth} to derive contributing terms for the 5-point MHV one-loop amplitude, while neglecting unphysical contributions:  tadpoles,  $(++)$ or $(--)$ bubbles due to anticipated cancellation by counterterms, and  $(+-)$ bubbles on external legs as these would yield a tree-level contribution upon amputation. Even then, the number of contributions from Eq.~\eqref{eq:olea_zth} exceeds those from Eq.~\eqref{eq:OLEA_Z}, indicating that the latter is indeed more efficient for computing higher multiplicity amplitudes.

\section{Acknowledgments}
\label{sec:ack}
H.K. is supported by the National Science Centre, Poland grant no. 2021/41/N/ST2/02956. P.K. is supported by the National Science Centre, Poland grant no. 2018/31/D/ST2/02731.

\bibliographystyle{JHEP}
\justifying
\bibliography{sample}
\end{document}